\documentstyle[12pt,aasms4,tighten,epsf]{article}
\def\W50{W$_{50}$~}
\journalid{337}{}

\lefthead{Graham, Clowes \& Campusano}
\righthead{The completeness of quasar surveys}

\begin{document}
\hskip 3.5in{Version 1.1 \hskip 10pt 3 December 1997}
\title{A new assessment of the completeness of quasar surveys: implications
for the luminosity function}

\author{Matthew J. Graham, Roger G. Clowes} 
\affil{Centre for Astrophysics, University  of Central Lancashire, 
Preston PR1 2HE, U.K.; m.j.graham@uclan.ac.uk; r.g.clowes@uclan.ac.uk}
\and
\author{Luis E. Campusano} 
\affil{Observatorio Astron\'{o}mico Cerro Cal\'{a}n, Departamento de
Astronom\'{\i}a, Universidad de Chile, Casilla 36-D, Santiago, Chile;
luis@das.uchile.cl}

\begin{abstract}

We apply a simple statistical method (Derenzo \& Hildebrand 1969) to
estimating the completeness of quasar surveys. It requires that an area has
been covered by two or more, preferably different, selection techniques. We
use three suitable data sets with separate selections from: variability and
UV-excess (170 quasars); objective prism and UV-excess (141 quasars);
multicolour and X-ray ({\it ROSAT,} 19 quasars). We find that, for selection
by UV-excess, the common limit of $U-B \le -0.35 \pm -0.05$ leads to losses
of $\sim 35\%$, typically missing low-luminosity $(M_{B} \gtrsim -24.5)$
quasars, independently of redshift. Systematic incompleteness will therefore
affect the new generation of large quasar surveys that select by $U-B \le
-0.35$. By correcting for this incompleteness, we find, from the first data
set ($B < 21.0$ and $z < 2.2$), that the evolution of the quasar luminosity
function (LF) is best described by joint luminosity and density evolution.
When extrapolated to $z = 0$, the LF matches that of local Seyfert galaxies
better than any previous determination. The LF shows an increase in the
number of low-luminosity quasars at low redshifts and of brighter quasars at
intermediate redshifts, relative to the LF of Boyle et al. (1990). This
result is consistent with models in which quasars fade from an initial
bright phase.

\end{abstract}


\section{INTRODUCTION}
 
The completeness of a quasar survey should properly be specified by the
survey selection function, $P(M,z,SED)$, which gives the probability of
detecting a quasar as a function of absolute magnitude, redshift and
spectral energy distribution $(SED)$. The number of detected quasars in each
cell of $(M,z,SED)$ space is divided by $P(M,z,SED)$ to give the number of
quasars that could be found. Summation over all cells gives the total
population of quasars within the survey limits. In practice, $P(M,z,SED)$
can not usually be determined reliably, and many cells may be empty or too
poorly occupied for reliable estimates of the true number.

In this paper, we apply a statistical method to estimating the completeness
of quasar surveys. It requires that an area of sky has been covered by two
or more, preferably different, selection techniques. It does not require
$P(M,z,SED)$.

We adopt $H_{0}=50$ km s$^{-1}$ Mpc$^{-1}$, $q_{0}=0.5$ and
$\alpha=0.5$ ($f_{\nu} \propto \nu^{-\alpha}$).

\section{ESTIMATING COMPLETENESS: THE METHOD}

The method for estimating the completeness of a quasar survey involves the
comparison of the number of detected quasars and an estimate of the total
population. The total population is estimated using the method devised by
\cite{dh} for estimating scanning efficiencies in the detection of events in
particle physics. The use of the Derenzo \& Hildebrand (subsequently DH)
method here is appropriate because in both applications the size of a total
set of events is estimated from repeated scans of the set, with each scan
yielding only a subset. In DH, the total number of particle events is
estimated from repeated scans of film records; in this paper, the total
number of quasar events is estimated from repeated scans (surveys) of an
area of sky.

In a similar application to ours, Harwit \& Hildebrand (1986) used the DH
method to estimate the remaining number of observational phenomena still to
be discovered in astronomy. Four kinds of scans were defined according to
their usage of (i) optical continuum methods; (ii) radio techniques; (iii)
optical spectroscopy; and (iv) all other techniques to discover astronomical
phenomena.

As a demonstration, we have applied the method to two samples of objects
derived from a known parent population. We show that the total number of
objects in the parent population is satisfactorily recovered.

The net visibility, $v$, of an object is defined to be the probability that
it will be found by an {\em average} detection technique. It takes values in 
the range 0 (invisible) to 1 (unmissable). Note that the net visibility
depends on both the intrinsic characteristics of the object and the
perceptiveness of the detection techniques.

The net visibility function, $F(v)$, is defined such that
$F(v) {\mathrm d}v$ is the number of objects with net visibilities in the
range $v$ to $v+{\mathrm d} v$. The total number of objects is then

\[ N_{T} = \int_{0}^{1} F(v) {\mathrm d} v \]

\noindent

Note that $F(v)$ is an {\em average} of visibility functions for the
individual techniques (Derenzo \& Hildebrand 1969). The existence of
visibilities and visibility functions is an assumption of the method. See
the Appendix for a formal derivation of $F(v)$ and the above result for
$N_T$.

If all objects had equal prior probability of being detected then $F(v)$
would be a $\delta$-function at the average detection probability. In
general, however, $F(v)$ is an unknown function that must be estimated.

Consider a population that is surveyed using several techniques, preferably
different but not necessarily. Suppose, for example, that there are three
techniques. A first technique ({\bf A}) detects $A$ objects, another
technique ({\bf B}) detects $B$ objects, and a third technique ({\bf C})
detects $C$ objects. A particular object may be detected by more than one
technique.

The average number of objects detected by one application of the average
technique, $M[1]$, is the average number detected by the three techniques

\[ M[1] = [A+B+C]/3 \]

\noindent
To determine the average number of new objects detected by the second
application of the average technique, we take the techniques in pairs. A new
object is detected if it is seen by one technique but not by the other.
Thus,

\[ M[2] = [Ab + Ac + Ba + Bc + Ca + Cb]/6 \]

\noindent
where $Ab$ denotes the number of objects detected by technique {\bf A} but
missed by technique {\bf B}, and so forth. Similarly, the average number of
new objects found by the third application of the average technique is

\[ M[3] = [Abc + Bac + Cab]/3 \] 

In general, the distributions of undetected objects after one, two and $i-1$
surveys are respectively $F(v)(1-v)$, $F(v)(1-v)^{2}$ and $F(v)(1-v)^{i-1}$.
Hence the number of new objects found in the $i^{\mathrm{th}}$ survey using
the average technique will be

\begin{equation}
\label{mneq}
M[i] = \int_{0}^{1} v F(v) (1-v)^{i-1} {\mathrm d} v 
\end{equation}

\noindent
The collection of values $M[1], M[2], \ldots, M[n]$ can then be used to
evaluate $F(v)$, if we assume for $F(v)$ an expression with not more than
$n$ parameters, where $n$ is the number of surveys.

\cite{hh} argue that, in general, the total number of objects, $N_{T}$,
derived from $F(v)$ will be approximately the same for any smooth trial
function, $F(v)$, whose $m$ parameters $(m \le n)$ are adjusted to give
satisfactory fits to the values of $M[i]$. \cite{dh} show that for any
function $F(v)$ that is integrable in the region $0 \le v \le 1$, and for
all values of $M[i], i=1, \ldots,\infty$, which satisfy equation \ref{mneq},
a lower limit for $N_{T}$ is given by

\[ N_{T} \ge \left[ \sum_{i}^{n} M[i] \right] + \left( \frac{M^{2}[n]}
{M[n-1]} \right) \left[ 1-\frac{M[n]}{M[n-1]} \right] ^{-1} \]

\noindent
subject only to statistical fluctuations in the measured quantities $M[1],
\ldots, M[n]$.

In the case where $n=2$, this expression can be rewritten as

\[ N_{T} \ge \frac{({\bf A} \cup {\bf B} + {\bf A} \cap {\bf B})^{2}}{4
({\bf A} \cap {\bf B})} \] 

\noindent
This shows that, when there are only two surveying techniques, the
estimated lower limit for $N_T$ depends on the number of objects that
both techniques have in common $({\bf A} \cap {\bf B})$. When the total
number of objects detected by both techniques is very large but the
number of objects in common is relatively small, this dependency will
lead to a very large estimate of $N_T$. From an analysis of the behaviour
of the above expression in these circumstances, we have found that
unreliable estimates of $N_T$ tend to be given when
$({\bf A} \cup {\bf B})/({\bf A} \cap {\bf B}) \gtrsim 5$. We therefore
recommend that the method is applied only to data sets in which this ratio is
$\lesssim 5$. There are no such discontinuities for higher values of $n$.

As a demonstration of the method, we consider the optical survey by Tritton
\& Morton (1984). They identified all objects in a region of 0.31 deg$^{2}$,
finding 747 ordinary stars, 3 white dwarfs, 4 quasars, and 143 galaxies. We
take the 601 stars found with $B \le 20.0$ as a complete sample and then
isolate two subsets by different selection criteria: (i) 379 stars with
$B-V<1.0$; (ii) 312 stars with an objective-prism classification of spectral
type earlier than $K$. Subset 1 has 111 stars not found in subset 2; subset
2 has 44 stars not found in subset 1. This gives $M[1]=345.5$ and
$M[2]=77.5$ and hence $N_{T} \ge 445$. The trial functions used in the
following sections give estimates of the total number in the range 445--667,
which embraces the actual total of 601 stars. The median estimate is 546.

\section{ESTIMATING COMPLETENESS: THE SURVEYS}

The method estimates the total number of quasars in the parameter space
defined by the boundaries of a data set --- that is, the total number of
quasars lying within the magnitude, redshift and SED ranges of the data set.
We have applied the method to three suitable data sets, each based on two
different survey techniques but satisfying our condition that $({\bf A} \cup
{\bf B})/({\bf A} \cap {\bf B}) \lesssim 5$.

{\it Data set 1\/} is based on a survey of quasars in ESO/SERC field 287 by
\cite{hv95} (see also: \cite{hv93}; \cite{vh95}). Quasar candidates were
selected primarily by variability ($s>3.5$, where $s$ is a variability
parameter typically ten times the amplitude of the variability), although a
UV-excess (UVX) criterion $(U-B < 0)$ was also used. The survey actually
consists of three samples: (i) a faint sample of 111 quasars covering 4.6
$\deg^{2}$ with $19.5 < B < 21.0$ and $z < 2.2$; (ii) a bright sample of 117
quasars covering 18.8 $\deg^{2}$ with $B < 19.5$ and $z < 2.2$; (iii) and a
high-redshift sample of 86 quasars covering 18.8 $\deg^{2}$ with $B < 21.0$
and $z > 2.2$. We construct data set 1 from the faint sample and those
quasars from the bright sample that lie within the boundaries of the faint
sample. It comprises 170 quasars covering 4.6 $\deg^{2}$ with $B < 21.0$ and
$z < 2.2$. We have not considered quasars with $z > 2.2$ as they were not
sampled by two independent methods but were selected jointly on their colour
and variability. The intrinsic luminosity limit for data set 1 is
$M_{B}=-19.83$.
145 quasars would be detected by their variability $(s>3.5)$
and 163 by their colour $(U-B < 0)$, giving $M[1]=154$. Only 7 quasars are
found by variability that are not found by UVX whereas UVX finds 25 quasars
not detected by their variability, which gives $M[2]=16$.

{\it Data set 2\/} is based on two quasar surveys of ESO/SERC field 927: (i)
a sparse-sampling survey of 118 quasars, selected by the Automated Quasar
Detection software (AQD; \cite{c86}; \cite{ccb}), covering $25 \deg^{2}$
with $B < 20.8$ and $z < 2.8$ (\cite{cc94}; \cite{g97}; Clowes, Campusano \&
Graham, in preparation); (ii) a survey of 151 quasars, selected by UV-excess
$U-B \le -0.3$, covering $14.4 \deg^{2}$ with $B < 20.0$ and $z < 2.8$
(\cite{g97}). Data set 2 is constructed from all the
UVX-selected quasars with $z < 2.2$ and all the AQD-selected quasars with $B
<  20, z < 2.2$ that lie within the boundaries of the UVX survey. It
comprises 141 quasars in $14.4 \deg^{2}$ and has an intrinsic luminosity
limit of $M_{B}=-21.04$.  
135 quasars were found in the UVX survey and
49 quasars in the AQD survey satisfying the relevant criteria. Only 6
quasars were found by AQD that were not detected by UVX whereas UVX found
92 quasars not detected by AQD. Thus, $M[1]=92$ and $M[2]=49$.

{\it Data set 3\/} is based on two surveys of Selected Area 57 (SA57): (i) a
survey of 30 quasars, selected by multicolour techniques, covering 0.29
deg$^{2}$ with $B < 22.6$ and $z < 3.12$ (\cite{kk}); (ii) a survey of 19
quasars, selected from {\em ROSAT\/} HRI observations of SA57, covering
$0.13 \deg^{2}$ with a limiting flux of $0.2 \times 10^{-14}$ erg s$^{-1}$
cm$^{-2}$ and $z < 3.02$ (\cite{mcsb}). Data set 3 is constructed from all
{\em ROSAT-}selected quasars and all quasars from Koo \& Kron that lie
within $12'$ of the {\em ROSAT} pointing centre. It comprises 19 quasars
covering $0.13 \deg^{2}$. 17 quasars were identified by Koo \& Kron that
meet the relevant criteria and 19 by {\em ROSAT}. No quasars were found by
Koo \& Kron that were not found by {\em ROSAT} and 2 were found by {\em
ROSAT} that were not found by Koo \& Kron. Thus, $M[1]=18$ and $M[2]=1$.
Note, however, that of the 33 sources detected by {\em ROSAT} in SA57, 7 are
classed as faint stellar objects. These could be quasars but they would be
fainter than the magnitude limit of $B_{J}=22.5$ of Koo \& Kron
(\cite{m97}).

We have used the values of $M[1]$ and $M[2]$ for the three data sets to fit
different trial functions, $F(v)$. The $F(v)$ and the corresponding
estimates of the total number of quasars are given in Table~\ref{result}.
The figure in parentheses after each estimate is the completeness of the
data set based on the estimated total number for the data set. The lower
limit to the total number of quasars is also given. With two techniques
contributing to a data set the trial functions have only two free
parameters. However, the trial functions give a good coverage of the
possible functional forms of the true visibility function.

The estimates show that: data set 1 is 63--99\% complete; data set 2 is
47--72\% complete; and data set 3 is 61--100\% complete. If we use the
median estimates then: data set 1 is 89\% complete; data set 2 is 54\%
complete; and data set 3 is 90\% complete. If we assume that there were
errors of $\sqrt N$ in the numbers of common $(A, B)$ and distinct $(Ab,
Ba)$ objects comprising each survey then the typical error in the estimate
of the total number of objects is $\sim$5\% in data set 1, $\sim$5\% in
data set 2, and $\sim$15\% in data set 3.

The success rates, for median estimates, of the individual selection
techniques are: variability (with $s>3.5$) selects 76\% and UVX (with $U-B <
0$) 86\% of the total population of quasars for data set 1; UVX (with $U-B
\le -0.3)$ selects 52\% and AQD 19\% of the total for data set 2; (the AQD
survey is by sparse sampling); colour selection finds 81\% and X-ray 90\% of
the total for data set 3.

\section{COMMENTS ABOUT SURVEYS} 

Random incompleteness --- effectively sparse sampling --- does not affect
the expectation values of any statistics if the sampling rate is known
(\cite{k86}). However, systematic incompleteness (e.g.\ bias against the
detection of low-redshift or low-luminosity quasars) means that a sample is
non-random and incorrect conclusions may be drawn.

Samples of quasars selected by UV-excess ($U-B \le -0.35 \pm 0.05$) are
normally assumed, following \cite{v83}, to contain at least 90\% of the
quasars with $z < 2.2$ within a given survey. However our results suggest
that a large fraction of quasars with $z < 2.2$, possibly as high as 50\%
(from data set 2), do not have a large UV-excess and must have been missed
by previous UVX surveys, e.g.\ \cite{bfsp}. Such losses could have serious
consequences if the incompleteness is systematic. We can test for systematic
incompleteness by examining the effectiveness of selection by UVX for
different $U-B$ limits. To do this, we take data set 1 and consider UVX
limits of $U-B \le -0.35$ and $U-B \le 0.0$.

One possibility is that a limit of $U-B \le -0.35$ is biased against low
redshift quasars $(z < 0.7)$. A plot of $U-B$ against redshift for all
quasars in data set 1 suggests that this is likely (see Hawkins \&
V\'{e}ron 1995, Fig.~1). For data set 1,
Table \ref{result} shows the effectiveness of selection by UVX for $z >
0.7$. The figure in parentheses after each estimate is the effectiveness of
selection by UVX with the relevant UVX limit. The lower limit to the total
number of quasars present is also given. If only low redshift quasars were
being missed for $U-B \le -0.35$ then for $z > 0.7$ the effectiveness for
the two UVX limits should be approximately the same. This is clearly not 
the case.

A second possibility is that a limit of $U-B \le -0.35$ is biased against
low luminosity quasars. Several authors have considered such a bias as
likely but estimate the incompleteness at $\sim 10$\% (e.g.\ Boyle et al.
1990). Note that a survey that arbitrarily defines a quasar to have $M_{B}
\le -23$ (Schmidt \& Green 1983) would also be biased against low luminosity
quasars. For data set 1, Table \ref{result} similarly shows the
effectiveness of selection by UVX for $M_{B} \le -24.5$. If only low
luminosity quasars were being missed for $U-B \le -0.35$ then for $M_{B} \le
-24.5$ the effectiveness for the two UVX limits should be approximately the
same. This is indeed the case.

% A third possibility is that a limit of $U-B \le -0.35$ is biased against a
% particular $SED$. For example, there could be a hidden population of
% dust-reddened quasars. Without the SEDs of the quasars in data set 1, we can
% not test for this type of bias. However, data set 3 suggests that there is
% not a significant hidden population.

In summary, a quasar sample selected by $U-B \le -0.35$ is systematically
incomplete, being biased in particular against low luminosity quasars. Part
of the incompleteness is probably caused by detectable radiation from the
host galaxy. This will redden the quasar and it will cause losses in surveys
that require candidates to be classified as star-like. 

From Parkes flat-spectrum quasars, \cite{fran} have suggested the existence
of three categories: (i) conventional blue quasars (40\%); (ii) red quasars,
with reddening from synchrotron radiation (30\%); (iii) very red quasars,
with reddening from dust (30\%). These last two categories of red quasars,
being faint in $B$, could be the type of objects that comprise the low
luminosity population suggested by our analysis.

The incompleteness revealed by this work, has implications for some of the
new generation of large quasar surveys. For illustration, consider the AAT
2dF quasar survey (\cite{sbscm}). This currently specifies $U-B \le -0.36$
and $18.25 < B < 21.0$, and has a projected final completeness of
$\sim 90\%$. Adapting data set 1 to have the same selection criteria gives
$M[1]=124$ and $M[2]=29.5$. The corresponding lower limit to the total
number of quasars is $N_{T} \ge 163$ and the median value of the five
estimators is 217. 104 quasars are detected with $U-B \le -0.36$. Thus, our
results imply that at best the AAT 2dF survey will select only 64\% of all
quasars with $18.25 < B < 21.0$ and $z < 2.2$.

\section{THE LUMINOSITY FUNCTION REVISITED}

Previous determinations of the quasar luminosity function (LF) could have
significantly underestimated the density of low-luminosity quasars. We can
now estimate the incompleteness, correct for it, and so determine the `true'
LF.

For data set 1, we have divided the ranges of apparent magnitude and
redshift into bins: $16 < B \le 18$, $18 < B \le 19$, $19 < B \le 20$, $20 <
B \le 21$; $0 < z \le 0.7$, $0.7 < z \le 1.5$, $1.5 < z \le 2.2$. Within
each joint magnitude and redshift bin, we have estimated the `true' surface
density of quasars using both the maximum completeness estimator,
$F(v)=K\delta(v-a)$, and the median completeness estimator,
$F(v)=K(1-v)^{a}$. We have generated `best-fit' models to these data using
maximum likelihood techniques for various parametric forms for the LF. We
consider two models involving pure luminosity evolution with parametric
forms given by \cite{bsp}: a smoothed double power-law function

\[ \Phi(M,z) = 
  \frac{\Phi^{\ast}}{10^{0.4(M-M(z))(\alpha + 1)} + 
10^{0.4(M-M(z))(\beta+1)}}  \]

\noindent
and a single power-law function

\[ \Phi(M,z) =   \frac{\Phi^{\ast}}{10^{0.4(M-M(z))(\alpha + 1)}}  \]

\noindent
where, in both cases, the redshift dependence, M(z), can be expressed either
as a $(1+z)$ power-law evolution, $M(z) = M^{\ast} - 2.5k_{L} \log(1+z)$, or
as an exponential evolution with look-back time, $M(z) = M^{\ast} - 1.08k_{L}
\tau$. The look-back time expressed as a fraction of the age of the
universe is $\tau(z) = 1-(1+z)^{-3/2}$. $\Phi^{\ast}$ is a normalisation
factor.

\noindent
We consider also a model with density and luminosity evolution parameterized
as

\[ \Phi(M,z) =
 \frac{\Phi^{\ast}\rho_{D}(z)}{10^{0.4(M-M(z))(\alpha + 1)} + 
10^{0.4(M-M(z))(\beta+1)}}  \]

\noindent
where the density evolution function is expressed as an exponential with
look-back time

\[ \rho_{D}(z) = \exp (k_{D}\tau) \]

\noindent
$\alpha$, $\beta$, $M^{\ast}$, $k_{L}$ and $k_{D}$ are the free parameters
that determine the fit.

The maximum-likelihood solution is found in each case by minimizing the
function:

\[ S = -2 \sum_{i,j} x_{ij} \log_{e} \mu_{ij} + 2 \sum_{i,j} \mu_{ij} \]

\noindent
where $x_{ij}$ is the observed surface density in magnitude bin
$i$ and redshift bin $j$, $\mu_{ij}$ is the expected surface density
in bin $i$ and bin $j$, and the summation is over all bins.

Each `best-fit' model was tested for goodness-of-fit using the Pearson
$\chi^{2}$ statistic. Table~\ref{bestfit} gives the best-fit parameters for
the models and their $\chi^{2}$ probabilities, $p(>\chi^{2})$. For
comparison, we have calculated how well some of the best-fit models of
\cite{bfsp}, normalised appropriately, fit the data; these are also included
in Table~\ref{bestfit}.

We can reject a single power-law LF and all of the best-fit models of Boyle
et al. Of the remaining models, all of which incorporate double power-law
LFs, the data are well described by those with density evolution. The best
fits are obtained when all evolution is expressed by exponentials with
look-back time. Fig.~1 shows our best-fit models applied to the data; for
comparison, the best-fit model of Boyle et al. (model B), normalised
appropriately, is also shown. At low redshifts the Boyle et al. model
underestimates the number of faint quasars and at intermediate redshifts it
underestimates the number of brighter quasars. Fig.~2 shows the LF
corresponding to our best-fit model extrapolated to $z=0$. For comparison
the figure also shows: the LF derived from local Seyfert galaxies of types 1
and 1.5 (Cheng et al. 1986); the LF for soft X-ray selected AGN
(\cite{ffcm}), extrapolated to $z = 0$; and Boyle et al. model B, extrapolated
to $z=0$. Our model provides the best-fit to the local Seyferts. The curve
for X-ray selected AGNs (including all types) lies a factor of $\sim 2$
above our LF, which suggests that other types of AGN (e.g. starburst
galaxies) are not related to quasars.

Compared with the most successful model of Boyle et al. (model B), our LF
shows an increase in the number of low luminosity quasars at low redshifts
and of brighter quasars at intermediate redshifts. It also matches well the
LF of local Seyfert galaxies. Such behaviour is consistent with quasar
models in which evolution is caused by the progressive exhaustion of the
fuel supply to the central black hole. Quasars fade from an initial bright
phase until a low, quasi-steady rate of energy production is reached. This
rate declines only slowly over a long timescale. Quasars in this final phase
of evolution are equivalent to local Seyfert galaxies.

\acknowledgments
The referee provided helpful comments. MJG acknowledges the support and
hospitality of Departimento d'Astronomia i Astrofisica at the Universitat de
Valencia. We thank Dr T. Miyaji for providing us with data before their
publication. LEC was partially supported by FONDECYT grant number 1970735.

\section*{APPENDIX}

In this appendix, we show that the total number of objects can be obtained from
an average visibility function.

Consider a data set that is scanned $N$ times. Each scan has its own
visibility function, $f_i(v)$. Let $X_i$ denote the number of objects found
by scan $i$ and $x_i$ be the number of objects not found by scan $i$ then, for
example, the expression $X_i x_j x_k$ represents the number of objects
found by scan $i$ but not by scan $j$ or scan $k$.

\parindent 0em

Consider the average number of objects found in one scan,

\[ M[1] = \frac{1}{N} \sum_{i}^{N} X_i \]

In terms of the visibility function for a scanner,

\[ X_i = \int_{0}^{1} v f_i(v) {\mathrm d}v \]

and so
\begin{eqnarray*}
M[1] & = & \frac{1}{N} \sum_{i}^{N} \int_{0}^{1} v f_i(v) {\mathrm d}v \\
     & = & \frac{1}{N} \int_{0}^{1} v \sum_{i}^{N} f_i(v) {\mathrm d}v 
\end{eqnarray*}

If we define the average visibility function by

\[ F(v) = \frac{1}{N} \sum_{i}^{N} f_i(v) \]

then 

\[ M[1] = \int_{0}^{1} v F(v) {\mathrm d}v \]

Now consider the average number of new objects found in the second scan,

\[ M[2] = \frac{1}{N(N-1)} \sum_{i,j(i \ne j)}^{N} X_i x_j \]

For a single scanner (fixed $i$),

\[ \frac{1}{N-1} \sum_{j (j \ne i)}^{N} X_i x_j = \int_{0}^{1} v f_i(v)
(1-v) {\mathrm d}v \]

and so
\begin{eqnarray*}
M[2] & = & \frac{1}{N} \sum_{i}^{N} \int_{0}^{1} v f_i(v) (1-v) {\mathrm d}v \\
     & = & \frac{1}{N} \int_{0}^{1} v \sum_{i}^{N} f_i(v) (1-v) {\mathrm d}v \\
     & = & \int_{0}^{1} v F(v) (1-v) {\mathrm d}v
\end{eqnarray*}

In general, the average number of new objects found in the $i^{\mathrm{th}}$ 
scan will be
\begin{eqnarray*}
 M[i] = & \\ 
 & \frac{1}{N (N-1) \ldots (N-i+1)} \sum_{i,j,\ldots,k 
(i \ne j \ne \ldots \ne k)}^{N} X_i x_j \ldots x_k 
\end{eqnarray*}

For a single scanner (fixed $i$),
\begin{eqnarray*}
\frac{1}{ (N-1)\ldots (N-i+1)} \sum_{j,\ldots,k (j \ne \ldots \ne k \ne i)}^{N} X_i x_j \ldots x_k \\ 
= \int_{0}^{1} v f_{i} (v) (1-v)^{i-1} {\mathrm d}v 
\end{eqnarray*}

and so 

\[ M[i] = \int_{0}^{1} v F(v) (1-v)^{i-1} {\mathrm d}v \]

Now consider the integral,
\begin{eqnarray*}
\int_{0}^{1} F(v) dv & = & \frac{1}{N} \int_{0}^{1} \sum_{i}^{N} f_i(v) 
{\mathrm d}v \\
                     & = & \frac{1}{N} \sum_{i}^{N} \int_{0}^{1} f_i(v) 
{\mathrm d}v \\
                     & = & N_T 
\end{eqnarray*}

Thus, fitting an average visibility function to the data allows the total
number of objects to be estimated.

\newpage

%\begin{figure}
%\epsfxsize=18.cm
%\epsfysize=18.cm
%\epsffile{fig1.ps}

\figcaption[fig1.ps]{This shows the `true' surface density of quasars
estimated in the various joint apparent magnitude and redshift bins using
the maximum (a,b) and median (c,d) completeness estimators. The top panel
for each estimated data set shows the predicted surface density using our 
best-fit LF model and the bottom panel that predicted by model B from Boyle
et al. (1990), normalised appropriately.}

\newpage

\figcaption[fig2.ps]{This shows the quasar LF based on our best-fit model
(solid line), extrapolated to $z=0$, and the LF for Seyfert galaxies (Cheng et
al. 1986) $(\ast)$. The dotted line indicates the extrapolation of the
quasar LF to magnitudes fainter than the limiting magnitude of the survey.
The dashed line is model B from Boyle et al. (1990), extrapolated to $z=0$,
and the dotted-dashed line is the LF for soft X-ray selected AGN
(Franceschini et al. 1994), extrapolated to $z=0$.}

\newpage

\begin{deluxetable}{llllllll}
\tablecolumns{8}
\scriptsize
\tablewidth{0pt}
\tablecaption{Data sets analysed and results. \label{result}}
\tablehead{
\colhead{} & \colhead{Data set 1} & \colhead{Data set 2} & 
\colhead{Data set 3} & \multicolumn{2}{c}{Data set 1: subset 1} &
\multicolumn{2}{c}{Data set 1: subset 2}\\
 & & & & \multicolumn{2}{c}{$(z > 0.7)$} & \multicolumn{2}{c}{$(M_{B} \le
-24.5)$}}
\startdata
Detection technique & UVX (0):163 & UVX (-0.3):135 &
Colour:17 & UVX (0):125 & UVX (-0.35):98 & UVX (0):60 & UVX
(-0.35):55 \nl
		    & Var.:145  & AQD:49 & X-ray:19  & Var.:117 &
Var.:117 & Var.:55 & Var.:55 \nl
$M[1]$              & 154             & 92 & 18 & 122.5 & 108.5 & 57.5 & 55 \nl
$M[2]$              & 16              & 49 & 1 & 8.5 & 18.5 & 4.5 & 5 \nl
No. of quasars   & 170             & 141 & 19 & 131 & 127 & 62 & 60 \nl
\tableline
Trial functions     & & \nl
$ F(v) =$ & & \nl
\tableline
$K(1-av)$,$v<a^{-1}$ & 258 (66\%) & 295 (48\%) & 29 (66\%) & 198 (63\%)
& 197 (50\%) & 94 (64 \%) & 91 (60\%) \nl
$0$,$v \ge a^{-1}$ & & \nl
$K(1-v)^{a}$ & 190 (89\%) & 302 (47\%) & 21 (90\%) & 141 (89\%) & 154
(64\%) & 67 (90\%) & 66 (83\%) \nl
$K \exp (-av^{2})$ & 271 (63\%) & 198 (71\%) & 31 (61\%) & 214 (58\%)
& 219 (45\%) & 100 (60\%) & 96 (57\%) \nl
$K \delta(v-a) $ & 172 (99\%) & 197 (72\%) & 19 (100\%) & 132 (95\%)
& 131 (75\%) & 62 (97\%) & 61 (90\%) \nl
$K , v<a $ & 171\tablenotemark{a} (99\%) & 262 (54\%) & 
18\tablenotemark{a} (\nodata) & 176 (71\%) & 175 (56\%) &
60\tablenotemark{a} (100\%) & 59\tablenotemark{a} (93\%) \nl
$0, v \ge a$ & & \nl
Lower limit  & 172 (99\%) & 197 (72\%) & 19  (100 \%) & 132 (95\%) & 131
(75\%) & 62 (97\%) & 61 (90\%) \nl
\enddata
\tablenotetext{a}{Solution has $a>1$; this is the value for $\int_{0}^{1} 
F(v) {\mathrm d} v$ (i.e. the upper limit is 1 and not $a$).}
\end{deluxetable}

\begin{deluxetable}{lclrlrlrlrlrlr}
\tablecolumns{14}
\scriptsize
\tablewidth{0pt}
\tablecaption{`Best-fit' parameters for evolutionary models derived using
maximum and median completeness estimators. \label{bestfit}}
\tablehead{
\colhead{Luminosity} & \colhead{Evolution} & \multicolumn{2}{c}{$\alpha$} & 
\multicolumn{2}{c}{$\beta$} & \multicolumn{2}{c}{$M^{\ast}$} & \multicolumn{2}{c}{$k_{L}$} &
\multicolumn{2}{c}{$k_{D}$} & \multicolumn{2}{c}{$P(\chi^{2})$} \\
\colhead{model} & & Max & Med & Max & Med & Max & Med & Max & Med & Max &
Med & Max & Med }
\startdata
2 power law & $(1+z)^{k_{L}}$ & -3.50 & -3.51 & -1.24 & -1.33 & -23.38 &
-23.60 & 2.29 & 2.20 & \nodata & \nodata & 0.8732 & 0.7087 \nl
2 power law & $e^{k_{L}\tau}$ & -3.56 & -3.56 & -1.26 & -1.35 & -22.26 &
-22.51 & 4.32 & 4.12 & \nodata & \nodata & 0.8572 & 0.6721 \nl
1 power law & $(1+z)^{k_{L}}$ & -1.83 & -1.83 & \nodata & \nodata & -22.71 &
-22.99 & 2.93 & 2.73 & \nodata & \nodata & 0.0048 & 0.0033 \nl
1 power law & $e^{k_{L}\tau}$ & -1.85 & -1.84 & \nodata & \nodata & -16.62 &
-21.39 & 5.15 & 4.72 & \nodata & \nodata & 0.0078 & 0.0040 \nl
2 power law & $(1+z)^{k_{L}}$ & -3.35 & -3.15 & -1.00 & -0.86 & -22.51 &
-21.97 & 2.95 & 3.36 & -1.06 & -1.79 & 0.9385 & 0.9064 \nl
2 power law & $e^{k_{L}\tau}$ & -3.42 & -3.33 & -1.07 & -1.00 & -21.23 &
-20.75 & 5.39 & 5.92 & -1.08 & -1.72 & 0.9392 & 0.9391 \nl
Boyle model B & $(1+z)^{k_{L}}$ & -3.87 & -3.87 & -1.32 & -1.32 & -22.37 &
-22.37 & 3.20 & 3.20 & \nodata & \nodata & 0.0410 & 0.0041 \nl
Boyle model M & $(1+z)^{k_{L}}$ & -3.90 & -3.90 & -1.34 & -1.34 & -22.59 &
-22.59 & 3.01 & 3.01 & 0.69 & 0.69 & 0.0162 & 0.0008 \nl
Boyle model N & $(1+z)^{k_{L}}$ & -3.89 & -3.89 & -1.29 & -1.29 & -22.79 &
-22.79 & 2.78 & 2.78 & 1.70 & 1.70 & 0.0006 & $4 \times 10^{-6}$ \nl
\enddata
\end{deluxetable}


\begin{thebibliography}{dum}
\bibitem[Boyle, Shanks \& Peterson (1988)]{bsp} Boyle, B. J., Shanks, T.,
\& Peterson, B. A. 1988, MNRAS, 235,935
\bibitem[Boyle et al. (1990)]{bfsp} Boyle, B. J., Fong, R., Shanks, T.,
\& Peterson, B. A. 1990, MNRAS, 243, 1
\bibitem[Cheng et al. 1986]{cdzf} Cheng, F. Z., Danese, L., de Zotti, G., 
Franceschini, A. 1986, MNRAS, 212, 857
\bibitem[Clowes 1986]{c86} Clowes, R. G. 1986, Mitt. Astron. Ges., 67, 174
\bibitem[Clowes \& Campusano 1994]{cc94} Clowes, R. G., \& Campusano, L. E.
1994, MNRAS, 266, 317
\bibitem[Clowes, Cooke \& Beard 1984]{ccb} Clowes, R. G., Cooke, J. A., \&
Beard, S. M. 1984, MNRAS, 207, 99
\bibitem[Derenzo \& Hildebrand (1969)]{dh} Derenzo, S. E., \& Hildebrand, R. H.
1969, Nucl. Instrum. Meth., 69, 287
\bibitem[Franceschini et al. 1994]{ffcm} Franceschini, A., La Franca, F.,
Cristiani, S., Martin-Mirones, J.M. 1994, MNRAS, 269, 683
\bibitem[Francis et al. (1997)]{fran} Francis, P., Webster R., Drinkwater
M., Masci F., \& Peterson B. 1997, AAO Newsletter 82, 4
\bibitem[Graham 1997]{g97} Graham, M. J. 1997, PhD Thesis, University of
Central Lancashire
\bibitem[Harwit \& Hildebrand (1986)]{hh} Harwit, M., \& Hildebrand, R. 1986,
Nat, 320, 724
\bibitem[Hawkins \& V\'{e}ron 1993]{hv93} Hawkins, M. R. S., \& V\'{e}ron, P. 
1993, MNRAS, 260, 202
\bibitem[Hawkins \& V\'{e}ron (1995)]{hv95} Hawkins, M. R. S., \& V\'{e}ron, P.
1995, MNRAS, 275, 1102
\bibitem[Kaiser 1986]{k86} Kaiser, N. 1986, MNRAS, 219, 785 
\bibitem[Koo \& Kron 1988]{kk} Koo, D. C., \& Kron, R. G. 1988, ApJ, 325, 92
\bibitem[Miyaji 1997]{m97} Miyaji, T. 1997, private communication
\bibitem[Miyaji et al. 1997]{mcsb} Miyaji, T., Connolly, A. J., Szalay, A. S., 
\& Boldt, E. 1997, A\&A, 323, L37
\bibitem[Schmidt \& Green 1983]{sg} Schmidt, M., \& Green, R. F. 1983, ApJ,
269, 352
\bibitem[Smith et al. 1996]{sbscm} Smith, R. J., Boyle, B. J., Shanks, T., 
Croom, S. M., Miller, L., \& Read, M. 1996, in Proc. IAU Symp. 179 New Horizons 
from Multiwavelength Sky Surveys (Reidel, Dordrecht), 348
\bibitem[Tritton \& Morton 1984]{tm} Tritton, K. P., \& Morton, D. C. 1984,
MNRAS, 209, 429
\bibitem[V\'{e}ron (1983)]{v83} V\'{e}ron, P. 1983, in Quasars and
Gravitational Lenses. (Univ. de Li\'{e}ge: Institut d'Astrophysique), 210
\bibitem[V\'{e}ron \& Hawkins 1995]{vh95} V\'{e}ron, P., \& Hawkins, M. R. S.
1995, A\&A, 296, 665

\end{thebibliography}
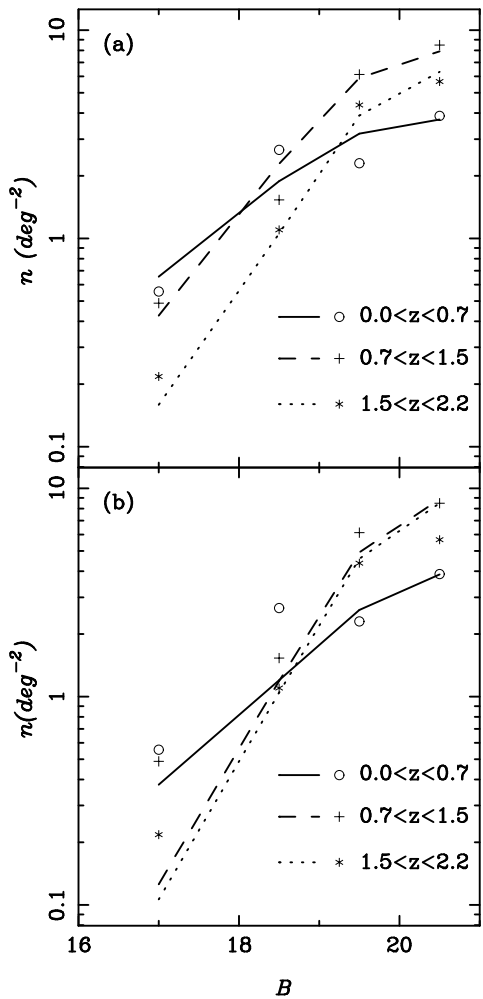
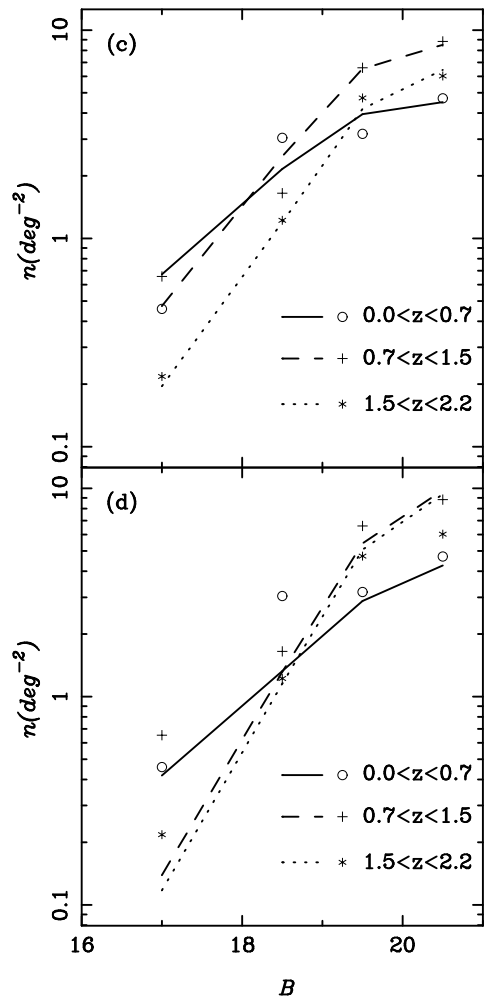
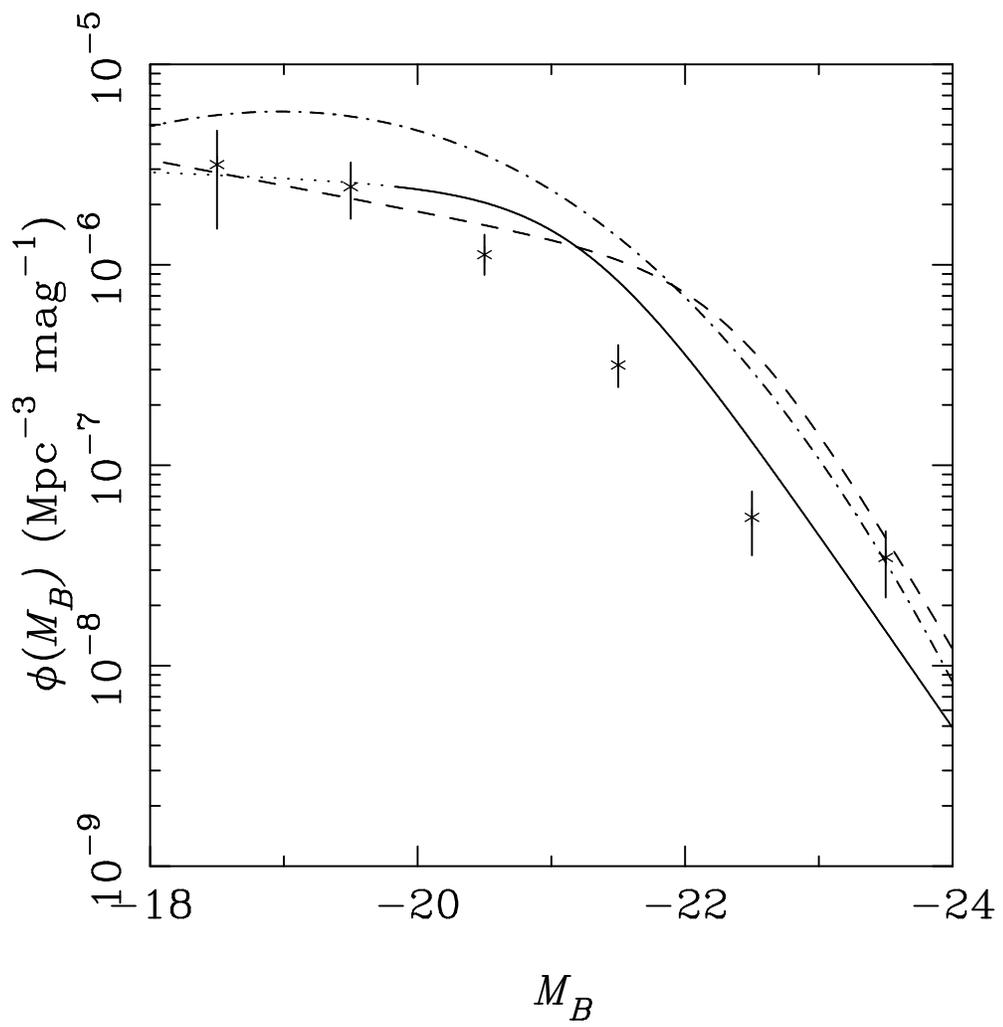
\end{document}